\documentclass[3p, twocolumn]{elsarticle}

\usepackage{hyperref}
\biboptions{sort&compress}

\usepackage{xcolor}

\begin{document}
\begin{frontmatter}

\title{DNA-functionalized gold nanoparticle assemblies for Surface Enhanced Raman Scattering}

\author[aff1]{D. Caprara}
\author[aff1]{F. Ripanti\corref{prova}}
\author[aff1]{A. Capocefalo}
\author[aff1,aff2]{A. Sarra}
\author[aff1]{F. Brasili}
\author[aff5]{C. Petrillo}
\author[aff5]{C. Fasolato\corref{prova2}}
\author[aff1]{P. Postorino\corref{prova3}}

\address[aff1]{Dipartimento di Fisica, Universit\`{a} Sapienza, P.le Aldo Moro 5, Roma, Italy}
\address[aff2]{Dipartimento di Scienze, Universit\`{a} di Roma Tre, Via Ostiense, Roma, Italy}
\address[aff5]{Dipartimento di Fisica e Geologia, Universit\`{a} degli Studi di Perugia, Via A. Pascoli, Perugia, Italy}

\cortext[prova]{Corresponding authors}
\ead{francesca.ripanti@uniroma1.it}
\ead{claudia.fasolato@unipg.it}
\ead{paolo.postorino@roma1.infn.it}


\begin{abstract}
The programmable assembly of DNA strands is a promising tool for building tailored bottom-up nanostructures. Here, we present a plasmonic nanosystem obtained by the base-pairing mediated aggregation of gold nanoparticles (NPs) which are separately functionalized with two different single-stranded DNA chains. Their controlled assembly is mediated by a complementary DNA "bridge" sequence. We monitor the formation of DNA assembled NP aggregates in solution, and we study their Surface Enhanced Raman Scattering (SERS) response by comparison with the single NP constituents. We interpret the revealed SERS signatures in terms of the molecular and NP organization at the nanoscale, demonstrating that the action of the DNA bridge molecule yields regular NP aggregates with controlled interparticle distance and reproducible SERS response. This demonstrates the potential of the present system as a stable, biocompatible, and recyclable SERS sensor. 
\end{abstract}


\end{frontmatter}


\section{Introduction}

In recent years, considerable efforts have brought to the development of novel strategies for realizing targeted nanostructures by bottom-up self-assembly processes \cite{bottomup1, bottomup2, bottomupDNA}. These represent a valid, cost-effective alternative to the established top-down nanofabrication techniques, as nanolithography or ion-beam milling \cite{nanolito1, ionbeam1}. In this framework, DNA-based nanomaterials have gained increasing attention, owing to the programmability of nucleic acid assembly: indeed, the pairing of complementary bases (adenine-thymine, cytosine-guanine) enables the realization of rationally-designed materials \cite{seeman2003dna, biffi2013phase, bomboi2016re, seeman2017, bomboi2019cold}. Among these novel nanotechnological platforms, hybrid systems composed of DNA and noble metal plasmonic nanostructures are promising candidates \cite{bottomup1, bottomupDNA}. Indeed, the possibility of coupling the self-assembly capabilities and thermal reversibility of DNA with the optical properties of metal nanoparticles (NPs) enables the realization of systems of interest for sensing applications based on Surface Plasmon Resonance (SPR) \cite{SPR-DNA, SPR-GG} or on Surface Enhanced Raman Scattering (SERS) \cite{kneipp1998detection, guerrini2015}. 

A plasmonic NP, when illuminated by light in resonance conditions with the collective oscillation of the electron gas of the metal, acts as a nanoantenna, focusing intense electromagnetic fields at the NP surface, far beyond the diffraction limit \cite{plasmonic}. This effect can be exploited for the ultrasensitive optical spectroscopy of molecules bound/located close to the NP surface \cite{colloids2016}. SERS spectroscopy has enabled the detection of analytes down to the femtomolar concentration range, by coupling the molecular-specific vibrational fingerprinting capability with the enhanced sensitivity of the plasmonic substrate \cite{kneipp1998detection, domenici2011}. 

In some cases, the DNA-based SERS sensing schemes are aimed at revealing or investigating the DNA molecules themselves: recent works have demonstrated the ability to determine the composition of DNA strands \cite{binren2015,guerrini2015} and the DNA-metal hybridization/conjugation dynamics \cite{guerrini2016}. Single base substitutions have also been detected by SERS in single-stranded DNA chains \cite{papadopoulou2011, belldick2017}. In other cases, DNA has been used as active element modulating the plasmonic efficiency of the substrate. In these applications, DNA is not the molecule directly studied by SERS, but its presence or its conformational changes trigger the SERS activity via the formation of hot-spots between NPs functionalized with standard and efficient SERS markers \cite{picciolini2014,aptamer1, aptamer2, aptamer-moskovits}. 

In order to promote the localization of single/double strands of DNA on the metal NP surface, overcoming the electrostatic repulsion between the negative charges of DNA and NPs, salts as MgSO$_4$ are commonly employed \cite{bell2006surface,binren2015}. Another possibility is the \textit{ad hoc} synthesis of metal NPs stabilized by a positive ligand \cite{guerrini2015}. Indeed, the immobilization of the DNA strands on the metal surface is most commonly obtained stimulating the electrostatic attraction of the DNA by positively charged/stabilized colloidal NPs. It is otherwise possible to purchase properly designed, thiolated DNA strands and induce their covalent binding to the NP surface. 

The covalent conjugation can be preferred in a series of applications, particularly those benefiting from a more stable linkage between the components of the nanosystem upon variations of temperature and pH of the environment. A stable conjugation is to prefer also in the perspective of designing a recyclable SERS substrate, where the analytes are typically released by chemical or physical methods after being trapped and revealed within the hot-spots of the structure \cite{recycle1, recycle2}. The coverage of the NP surface by covalently immobilized ligands allows the control of the composition of the NP environment also in complex media, preventing the accidental adsorption of molecules on the NP surface, within the highly sensitive region for SERS and SPR experiments. It is worth mentioning that a covalent DNA-NP conjugation at a specific molecular concentration grants the stable and predictable orientation of the DNA strands onto the metal surface \cite{doi:10.1021/ja972332i}. This is favorable for inducing the DNA pairing and assembly. The controlled configuration is also very important in the SERS response of the system, since the SERS spectrum of DNA strands is known to be strongly affected by the molecular orientation on the metal surface \cite{barhoumi2008, papadopoulou2011}.
 
Here, we report the study of gold (Au) NPs functionalized with two single-stranded DNA sequences. The SERS signature was characterized for the two chains by comparison with their Raman spectra to identify the spectral markers of the DNA scaffold. The self-assembly of these two hybrid DNA-NP systems within aggregates was then achieved by the action of a "bridge" molecule, in halves complementary to each of the sequences attached to the NPs. The DNA bridge acts therefore as a molecular glue. The temperature dependent UV-Visible spectra acquired on the NP aggregates during their formation demonstrate that the aggregation is specific and DNA mediated. Although a specific spectroscopic marker of the DNA pairing (\textit{e.g.} a band ascribed to hygrogen-bonding) was not detected, a change in the system spectroscopic signature arising from the bridge molecules could be identified in the aggregated state. Notably, our study is conducted with gold NPs, while the vast majority of SERS studies on DNA rely on silver substrates \cite{binren2015, guerrini2016, marotta2013limitations}, with rare exceptions \cite{Thacker2014DNAOB}. Silver nanosubstrates are in general more efficient for the SERS amplification, but poorer from the point of view of the chemical stability and biocompatibility compared to the gold counterpart \cite{fratoddi, toxicity}. 

The SERS study of DNA here reported is sequence-specific, reflecting the nucleotide composition of the DNA strands used, and allows monitoring the occurrence of the NP aggregation, driven at controlled interparticle distance by the self-assembly of DNA. The comparison between SERS spectroscopy and electron microscopy measurements demonstrates the negative correlation between the SERS signal-to-noise ratio and the interparticle distance, suggesting a path for SERS signal optimization. 

Conveniently, the SERS bands of DNA are outside the fingerprint region of common analytes detected by SERS spectroscopy \cite{colloids2016, heck2018, capocefalo2019}. In perspective, this will allow the use of this system as a robust and controlled sensing platform. Indeed, the DNA mediated assembly of the plasmonic nanoarchitecture can be induced in presence of small analytes to detect, by trapping them within the NP aggregate and reversibly releasing them by ramping up the temperature. 

\section{Experimental}

\paragraph{Materials and characterization}
Stock solutions of 60 nm diameter AuNPs stabilized by citrate capping were purchased from Ted Pella, inc. (Redding, CA). Synthetic monophosphate DNA sequences with DUAL HPLC purification were purchased from Integrated DNA Technologies (IDT, USA). For the experimental realization of DNA-linked NP assemblies we used three different and purposely-programmed single-stranded DNA chains, selected according to the seminal work of J.J. Storhoff and coworkers \cite{storhoff}. We refer to them as chain-A, chain-B, and bridge. Chain-A and chain-B are respectively a $5'-$ and $3'-$ alkylthiol-capped 12-base oligonucleotides. They were used for the functionalization of the NPs. The bridge is a non-thiolated 24-base oligomer, complementary to the chain-A and chain-B sequences. It acts as a linker for the aggregate formation. The specific sequences are \\

\noindent
chain-A: \small{$5'$ X-S=S-(CH$_2$)$_6$-CGCATTCAGGAT $3'$} \\
chain-B: \small{$3'$ X-S=S-(CH$_2$)$_3$-ATGCTCAACTCT $5'$}\\
bridge: \small{$5'$ TACGAGTTGAGAATCCTGAATGCG $3'$}\\

\noindent
where the X-S=S- complex specifies the thiol protecting group. Indeed, due to the high reactivity of thiol groups, during the oligonucleotide synthesis, alkylthiol-capped DNA sequences are protected by a disulfide bond; a chemical reduction is necessary for ensuring the exposition of the thiol group for the NP functionalization. The preliminary Raman characterization was performed on the pristine molecular samples. Small volumes of the three different strands in aqueous solution were dropcasted on a gold leaf substrate and dried at room temperature.  

\paragraph{Nanoparticle functionalization}
The three DNA sequences were separately diluted in deionized Milli-Q water. The actual strand concentration in these solutions was determined using a NanoDrop Microvolume Spectrophotometer (ThermoFisher Scientific, USA). To deprotect the thiolated sequences (chain-A and chain-B) from the disulfide bonds, we dissolved 5 nmol of each strand in a 0.1 M Dithiothreitol (DTT) solution, allowing the reaction to continue for 2 hours. The procedure was followed by the purification through NAP-5 column (GE Healthcare UK). AuNP functionalization with the thiol-ending nucleotides was carried out according to the well-defined assay procedure reported in Ref. \cite{Hill2006TheBA}. Briefly, we initially prepared two different DNA-NP solutions, by separately adding an excess of thiolated DNA strands to a controlled volume of 60 nm AuNP stock solution. The NP dispersion was previously dialyzed for 24 hours to remove part of the citrate buffer, thus making the NP surface cleaner and available for the DNA attachment. Phosphate buffer (10 mM) and SDS surfactant (1$\%$) solutions were then added to the dispersion, for improving the wettability of the NP surface. A slow NaCl salting aging to a final 0.7 M concentration \cite{Hill2009TheRR} (done in 1 day) concluded the functionalization process. The salting aging is a critical and fundamental step that allows an optimal and homogeneous coverage of the NP surface, as it minimizes the electrostatic repulsion between the negative charges on both the oligonucleotide skeleton and the NP surface. Finally, to remove the unbound DNA strands, we centrifuged the dispersions at $15^{\circ}\,$C and $10000\,$rpm for $30$ minutes, removing the supernatant and finally adding a 0.3 M NaCl solution \cite{doi:10.1021/ja972332i}. The purification step was repeated five times. 
For SERS and Field Effect Scanning Electron Microscopy (FESEM) measurements on the single-chain functionalized NPs, the purified DNA-NP solution was deposited on a glass slide/silicon substrate and dried at room temperature.

\paragraph{DNA-NP controlled aggregation}
The bridge sequence mediated aggregation of the DNA-NP complexes was induced in solution and monitored by UV-Visible spectroscopy. We prepared two equimolar solutions of AuNPs functionalized with chain-A and chain-B respectively. They were mixed in the presence of an excess of complementary bridge molecules (the bridge concentration exceeded by four orders of magnitude the NP concentration, to guarantee the formation of large aggregates). The final solution was incubated for 20 minutes in a pre-heated Memmert oven at 75$^{\circ}$C, then slowly cooled down to room temperature overnight. To remove the fraction of unbound bridges the solution was centrifuged at 4400 rpm for 30 minutes. Subsequently, the supernatant was removed and the sample re-dispersed in a 0.3 M NaCl solution. The purification step was repeated three times. For SERS and FESEM measurements, the aggregated NP purified dispersion was deposited on the substrate and dried at room temperature.

\paragraph{UV-Visible spectroscopy and DNA aggregation measurements}
UV-Visible absorption measurements were carried out from Jasco v-570 double ray spectrophotometer equipped with a Jasco ETC-505T Peltier thermostat which allowed to perform measurements at controlled temperature. 
To monitor the DNA aggregation as a function of temperature, a volume of $\sim 250\,\mu$l of the mixed functionalized NPs and bridge solution was sealed in a quartz cuvette (optical path 1 mm). The system was heated up to $75^{\circ}\,$C, well above the melting temperature of the DNA strands, calculated to be $\sim 54^{\circ}\,$C using NUPACK Oligo-simulator \cite{nupack}. This condition ensured every sequence to be unbound from the others. The UV-Visible spectra were collected ramping down the temperature to $20^{\circ}\,$C in steps of $5^{\circ}\,$C (the thermalization time was 20 minutes). The cuvette was stirred before each acquisition, to ensure the sample homogeneity and to prevent the formation of precipitates. 

\paragraph{Raman/SERS spectroscopy}
Raman and SERS measurements were carried out using a Horiba HR-Evolution microspectrometer in backscattering geometry, equipped with a He-Ne laser, $\lambda = 632.8$ nm and 25 mW output power ($\sim 10$ mW at the sample). Raman spectra were acquired by illuminating the molecular depositions with full laser power. Using proper optical filters, SERS spectra were collected keeping the laser power below 0.25 mW to avoid laser heating and sample degradation, induced by the enhanced field intensity occurring in presence of the plasmonic NPs.
The detector was a Peltier-cooled charge-coupled device (CCD) and the resolution was better than 3 cm$^{-1}$ thanks to a 600 grooves/mm grating with 800 mm focal length. The spectrometer was coupled with a confocal microscope. A 100$\times$ (N.A. = 0.9) Olympus objective lens was used for the experimental acquisitions. Further details on the experimental apparatus can be found in Ref. \cite{colloids2016}.
The Raman and SERS spectra are presented after polynomial baseline subtraction. The SERS data are presented after a smooth processing (Savitzky-Golay method, 11 and 20 point windows for the single-chain and bridge mediated aggregates, respectively). A comparison between smoothed and raw data is presented in Section F of the Supplementary Material. 

\paragraph{Electron microscopy}
Samples for FESEM imaging were prepared as described above. 
Images were recorded using a Zeiss Auriga 405 microscope at Sapienza Nanoscience \& Nanotechnology Laboratories (SNN-Lab) of the Research Center on Nanotechnology Applied to Engineering (CNIS) of Sapienza University.
The acquired frames were analyzed by Gwyddion software \cite{gwyddion}, version 2.53, for characterizing the morphology of the SERS substrates in terms of AuNPs size and interparticle distance within clusters.
The detailed description of the analysis procedure by employing the radial autocorrelation function of the scattering intensity is reported in Section H of the Supplementary Material.

\section{Results and Discussion}

\begin{figure}[]
\centering
\includegraphics[width=7.8cm]{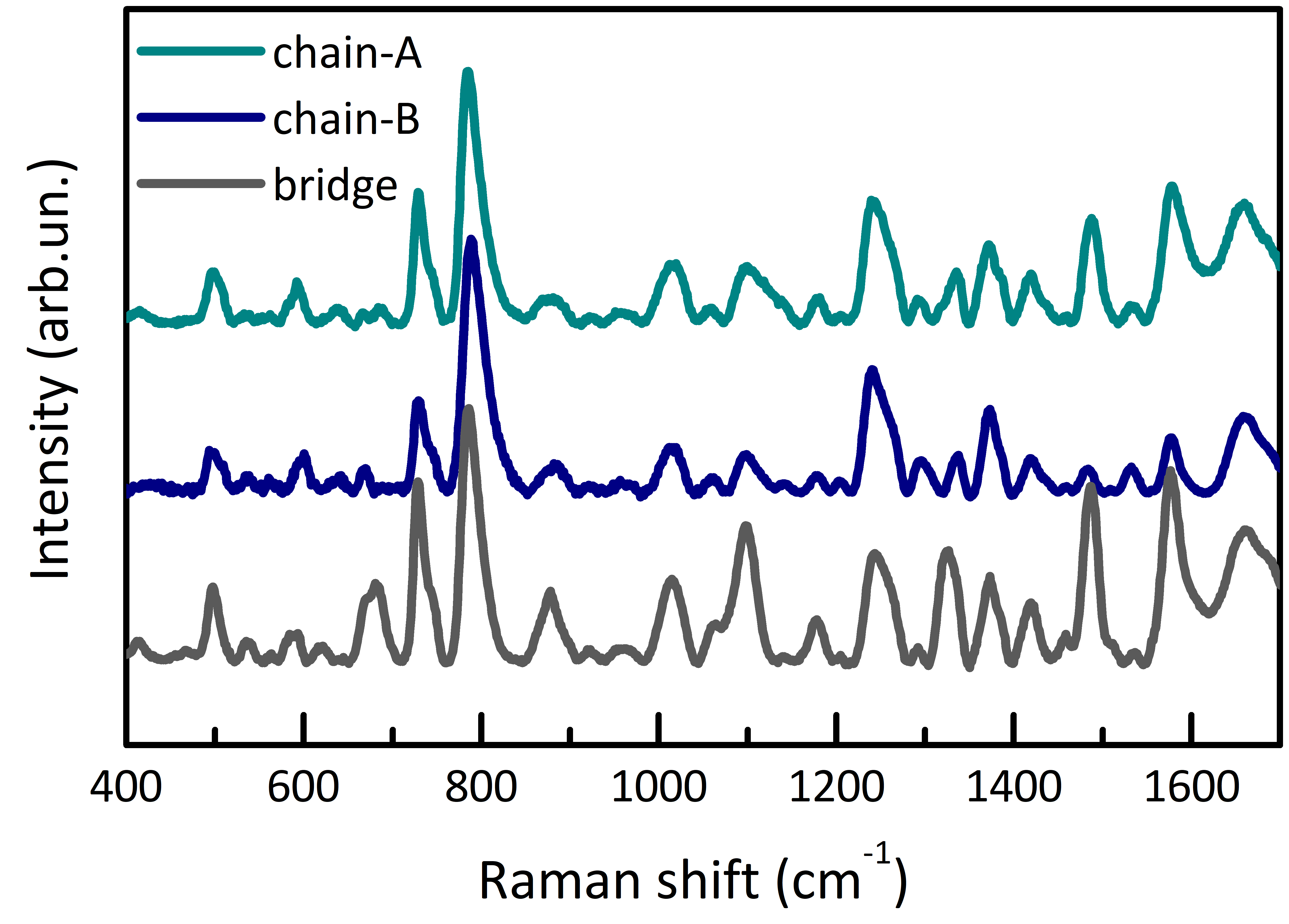}
\caption{\footnotesize Raman spectra of the single monophosphate chains: chain-A (cyan), chain-B (blue), and bridge (gray). } \label{Raman_characterization}
\end{figure}

\paragraph{Raman study of the DNA strands} Preliminary Raman measurements allowed to identify the spectral fingerprint of the selected DNA sequences. In Figure \ref{Raman_characterization} we show the spectra collected as described in the Experimental section. The spectra are normalized to the most intense peak, at $\sim 789$ cm$^{-1}$, associated to the convolution of the vibrational modes of the cytosine and of the O-P-O phosphate groups. In Figure S1 of the Supplementary Material, we show the Raman spectra of single triphosphate bases as a reference. Comparing the collected spectra with those of single bases provides a precise and complete assignment of the Raman peaks, as reported in the Supplementary Material (Table T1). 

The effect of the reduction step on the chains, \textit{i.e.} the transformation of the -S=S into -SH bond in chain-A and chain-B, can be studied by Raman spectroscopy. To do so, we measured the Raman signal before and after the cleavage of the disulfide protecting groups. We focused on the region around $\sim 2550$ cm$^{-1}$, where we expected to observe the -SH stretching Raman band. As shown in Figure S2, the -SH band is present in the spectrum of the chains only after the occurrence of the cleavage. This ensures that the chemical protocol here adopted is effective.  

\paragraph{DNA-AuNPs conjugation and SERS features} 

\begin{figure}[]
\centering
\includegraphics[width=7.8cm]{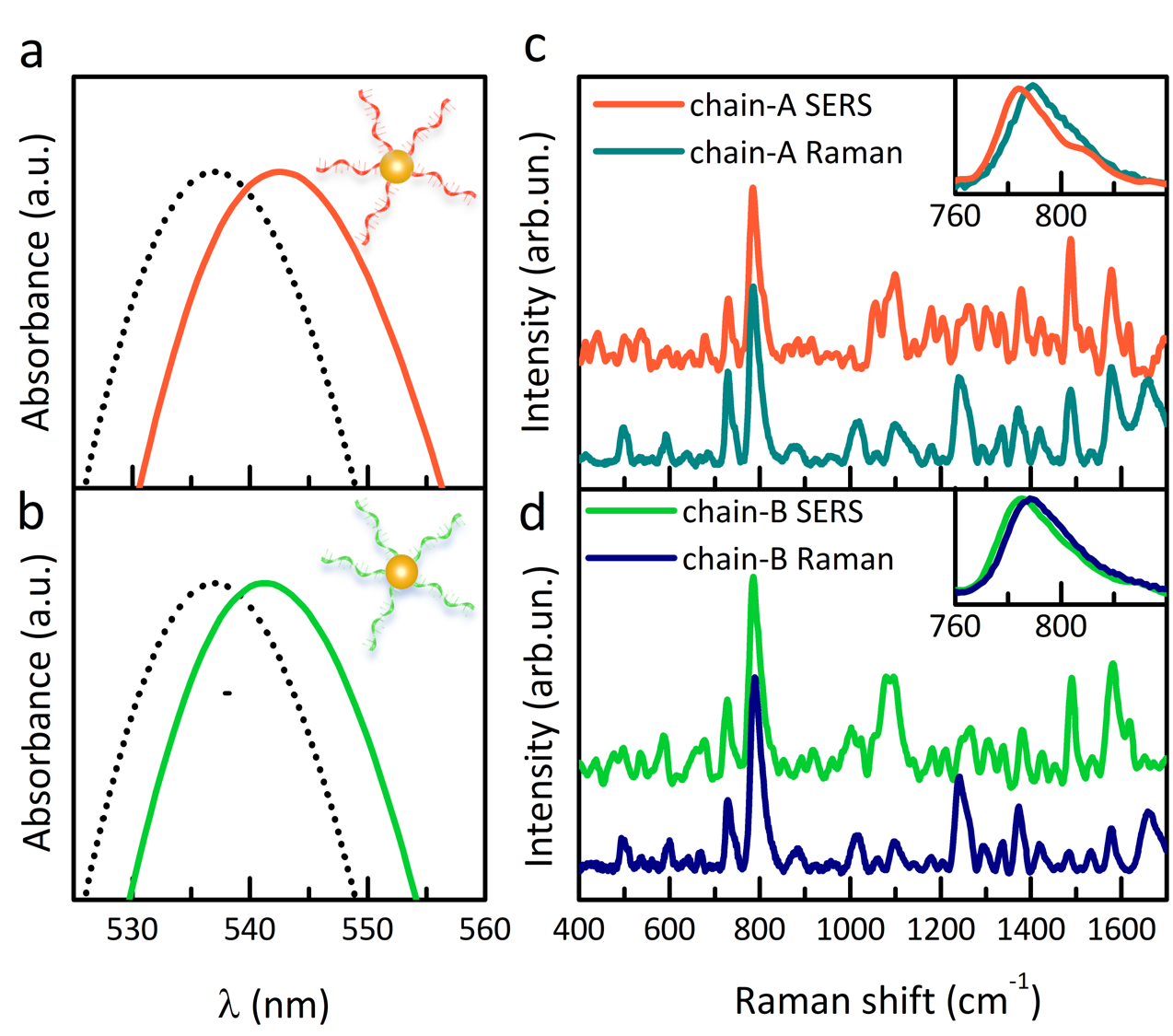}
\caption{\footnotesize Characterization of Au NPs functionalized with chain-A (top row) and chain-B (bottom row). Left column: zoom on the UV-Visible absorption spectra of bare, citrate capped AuNPs (dotted black curve in panels a and b) in the region of the plasmonic absorption peak are compared to the spectrum of AuNPs functionalized with the thiolated chain-A (orange, panel a) and chain-B (green, panel b). The corresponding full spectra are reported in Figure S3 of the Supplementary Material. Spectra are normalized to the maximum of the plasmonic peak. Right column: SERS spectra of functionalized AuNPs are compared to the Raman spectra of the corresponding DNA chain-A (panel c, orange: SERS; cyan: Raman) and chain-B (panel d, green: SERS; blue: Raman). The insets highlight the redshift of the most intense SERS band (cytosine + O-P-O) as compared to the Raman spectrum. This band was used for spectral normalization. } \label{SERS_single_chains}
\end{figure}

The DNA anchoring on the NP surface can be monitored by UV-Visible absorption measurements. In the UV-Visible spectrum of the pristine plasmonic NP dispersion, a well-defined peak ascribed to the NP surface plasmon resonance can be recognized. Its position depends on the size, shape and dielectric environment around the NP. In the case of spherical AuNPs, 60 nm diameter, dispersed in solution, the peak is centered at $\lambda = 535$ nm. When the NPs are functionalized with a ligand molecule, as DNA, the dielectric environment changes and a redshift of the plasmonic peak is observed \cite{capocefalo2019}. This is verified in the functionalization with both chain-A ($\Delta \lambda \sim 6$ nm) and chain-B ($\Delta \lambda \sim 4$ nm), as shown in Figure \ref{SERS_single_chains} a-b respectively, which ensures a successful and stable conjugation of the AuNPs with the DNA aptamers. The complete UV-Visible spectra for both the bare and the functionalized NP solutions are shown in Figure S3 of the Supplementary Material. 

The SERS features of DNA-AuNPs were investigated by carrying out direct spectroscopic measurements. The good agreement between the Raman and SERS spectra of chain-A and chain-B functionalized NPs is evident from their comparison in Figure \ref{SERS_single_chains} (panels c-d, respectively). The SERS spectral features in the region around $\sim 2550$ cm$^{-1}$, where the stretching vibration of the -SH group is expected, are shown in Figure S4 of the Supplementary Material. The lack of this contribution proves the absence of thiol compounds in the system, thus confirming the successful binding of the DNA molecules to the NP surface. Further, the optimized functionalization protocol minimizes the AuNP residual citrate capping, and this does not affect considerably the final SERS signal, as demonstrated by the reference spectrum in Figure S5. 

The SERS response is characterized by a good reproducibility of the spectral shape. 
However, a marked variability ($\sim100\%$) of the overall SERS intensity was retrieved during repeated acquisitions. It is known that most of the SERS signal revealed on NP-based SERS substrates originates from the hot-spots between aggregated NPs \cite{science2008, apl2014}. The intensity variability is here explained by the uncontrolled self-assembly of the NPs during the drying process, which yields an irregular hot-spot size distribution.

The Raman cross-section and typical SERS enhancement factors revealed for DNA are than other commonly adopted SERS analytes \cite{Thacker2014DNAOB}. This is an advantage for developing DNA-NP assemblies for SERS sensing, as the DNA signal would not cover that from other analytes, but it can be detrimental for obtaining structural information on the DNA scaffold. Nevertheless, despite the low intensity, the spectra in Figure \ref{SERS_single_chains} allow for the recognition of the principal SERS bands, which can be assigned thanks to the comparison with the Raman counterpart (see Table T1 in the Supplementary Material). Peaks ascribed to the four bases are identified, which provides a direct fingerprint of the DNA chemical composition. The spectra are dominated by an intense peak at $\sim 784$ cm$^{-1}$, assigned to the vibration of the cytosine and the O-P-O phosphate group components. As highlighted in the insets of Figure ~\ref{SERS_single_chains} c-d, the vibrational band observed in the Raman spectra at $\sim 789$ cm$^{-1}$ is redshifted in the SERS spectra for both the chains. The effect is typically observed in SERS spectra, being ascribed to the redistribution of the electronic cloud induced by molecular binding to the metal NP surface \cite{apl2014}.  
This band is used for the spectral normalization of both SERS and Raman data.

\begin{figure}[]
\centering
\includegraphics[width=7.8cm]{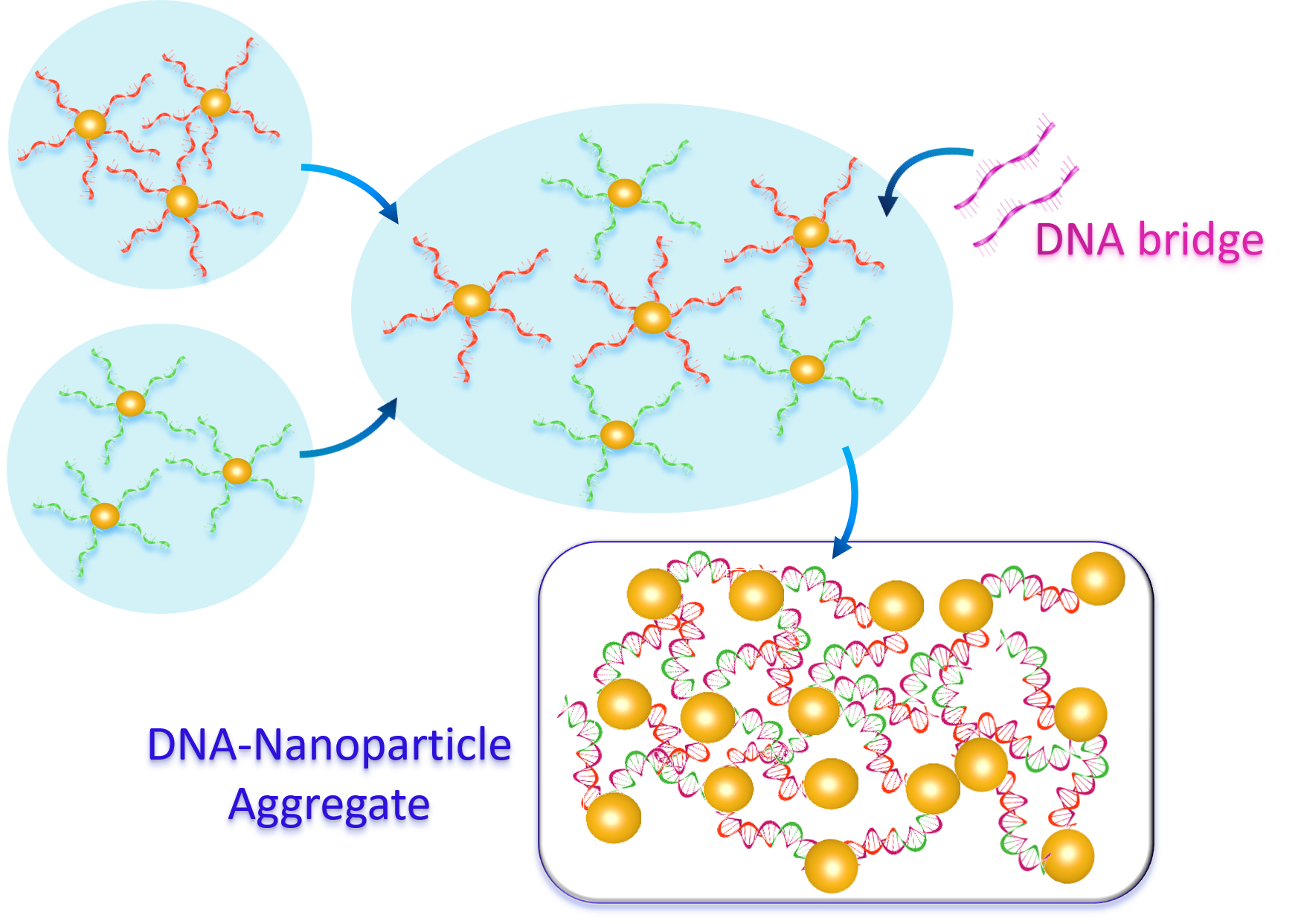}
\caption{\footnotesize Sketch of the implemented system: AuNPs, in water dispersion, are functionalized with two distinct 12-base DNA chains (-A and -B). The addition of a bridge sequence, complementary to the two chains -A and -B, induces the system hybridization, with the formation of mesoscopic aggregates.}
 \label{system}
\end{figure}

\paragraph{DNA-AuNp aggregates} Once the main components of our SERS nanosystems were characterized, we produced DNA-NP aggregates in which the bridge sequence acts as a selective linker between the functionalized NPs. The adopted experimental procedure adopted is described in the Experimental section, and it is sketched in Figure~\ref{system}. To directly monitor the NP aggregation as a function of temperature, which leads to the formation of mesoscopic NP assemblies, UV-Visible absorption measurements were performed as a function of temperature over the range 20-75$^o$C, starting far beyond the DNA melting temperature and slowly cooling down the system. Thermalization times were long enough to induce the gradual pairing of the DNA sequences and, consequently, the NP assembly. 

\begin{figure}[]
\centering
\includegraphics[width=7.8cm]{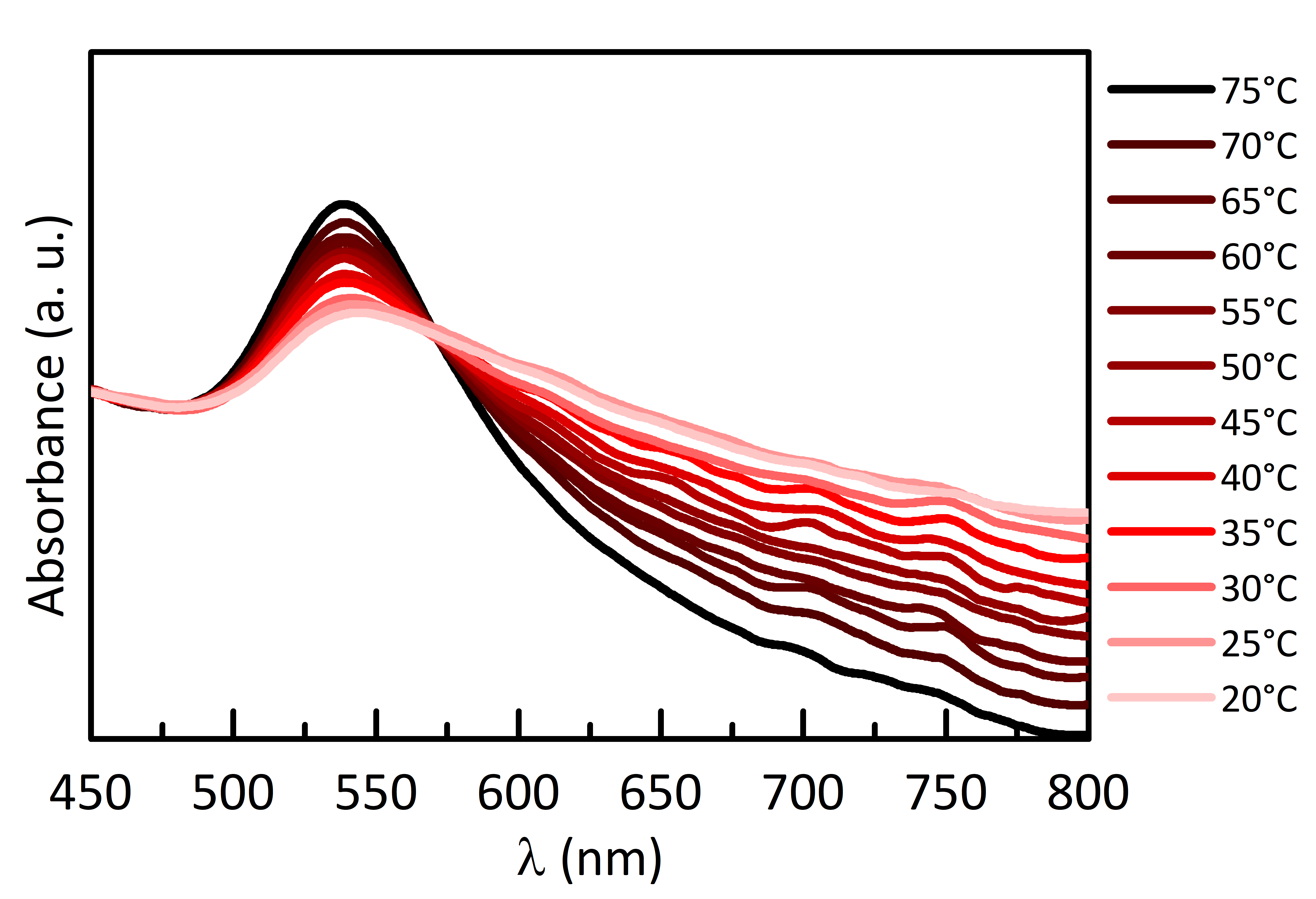}
\caption{\footnotesize UV-Visible absorption spectra acquired as a function of temperature on the hybrid system composed of the two AuNP dispersions (functionalized with chain-A and chain-B)  and the bridge molecules. As the temperature decreases, markers of the NP aggregation, as redshift and increasing peak width, become evident.} \label{UV_melting_SEM}
\end{figure}

In Figure \ref{UV_melting_SEM} we report the UV-Visible absorption spectra collected as a function of temperature during the NP aggregation by DNA pairing. At temperatures considerably higher than the system melting temperature, the UV-Visible profile can be associated to a dispersion of single, DNA-functionalized AuNPs (see Figure \ref{SERS_single_chains} a-b and Figure S3 in the Supplementary Material for comparison). As the temperature decreases, the complementary DNA chains pair up, forming aggregates with increasing size. The observed widening, redshift and decrease of the plasmonic absorption peak intensity, along with the increased extinction in correspondence of the red-NIR spectral region ($600 \div 800$ nm), are expected consequences of the formation of large aggregates \cite{capocefalo2019}. The temperature dependence of the observed behavior is a signature of the specificity of the DNA binding process, witnessing that the bridge mediated aggregation occurs by DNA hybridization into a double helix, which grants the NP assembly at controlled interparticle distance.

\begin{figure}[]
\centering
\includegraphics[width=7.8cm]{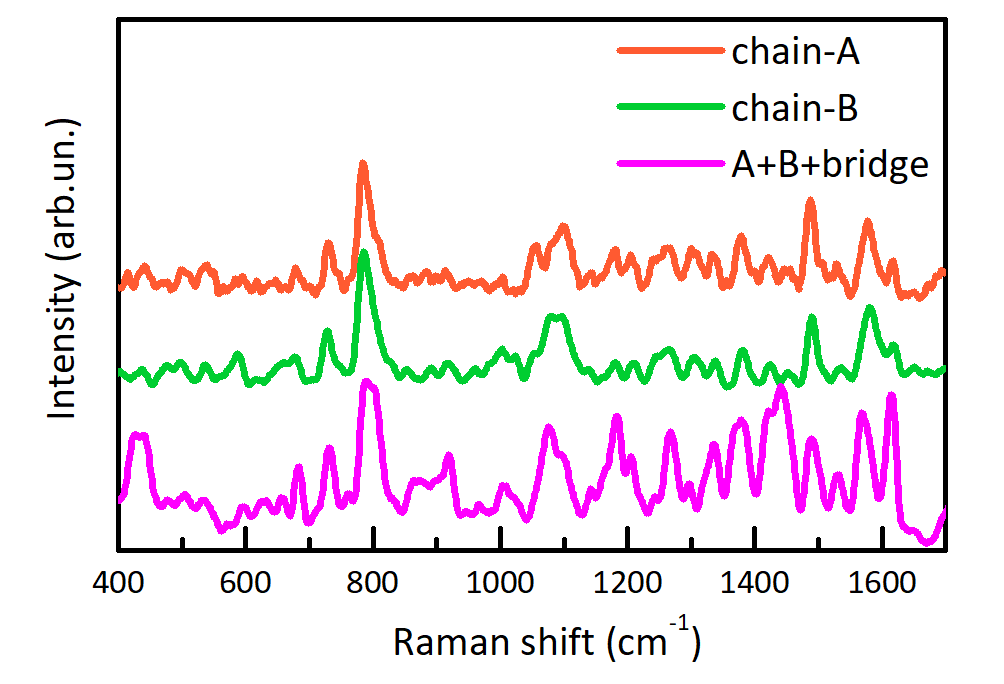}
\caption{\footnotesize SERS spectra of the single-chains, chain-A (orange) and chain-B (green), are compared with the SERS spectrum of the combined system A+B+bridge (magenta).} \label{SERS_comparison_all}
\end{figure}

As described, the SERS characterization of the aggregates was performed measuring a microvolume of the annealed DNA-NP solution dropcasted and dried on a glass slide. The SERS spectrum of the double-stranded DNA was measured, and in Figure \ref{SERS_comparison_all} we show the comparison between the SERS spectra of the two separated chains and that of the bridge-conjugated aggregates. Spectra are normalized to the most intense SERS band, $\sim 784$ cm$^{-1}$. 

The first information derived by the spectra in Figure \ref{SERS_comparison_all} is that no new peaks arise originating from the presence of the bridge molecule or to the hybridization are observed. This is not surprising since, in principle, spurious base-pairing can occurr accidentally also between non-complementary sequences as a consequence of the drying process. This partially prevents the spectroscopic discrimination between the specific hybridizations in the paired system and non-specific bindings, which can be observed also in the single-chain systems \cite{marotta2013limitations}.

Nevertheless, a remarkable spectral weight redistribution between the various components of the SERS signature can be observed in the spectrum of the bridge-conjugated system, which appears quite different from those of chain-A and chain-B functionalized NPs. These spectral modifications can be related to a partially different orientation of the chains with respect to the NP surface in the case of the conjugated system. Indeed, in the presence of the bridge molecule, we expect the DNA strands to stand upright onto the NP surface, because of the occurred interchain pairing. In the single-chain functionalized NPs, on the other hand, the molecules on the NP surface can move in a relatively more flexible manner, thus their orientation is not strictly determined. The orientation of the DNA molecules is known to critically affect the SERS spectral shape, mostly in the 1200-1600 cm$^{-1}$ region (high-frequency side of Figure \ref{SERS_comparison_all}) and in correspondence of the adenine band at 730 cm$^{-1}$ \cite{papadopoulou2011}. Fortunately, the use of thiolated molecules reduces this variability \cite{marotta2013limitations}, resulting in a less pronounced spectral modification in the present case compared to Refs. \cite{barhoumi2008, papadopoulou2011}. 

\begin{figure}[]
\centering
\includegraphics[width=7.8cm]{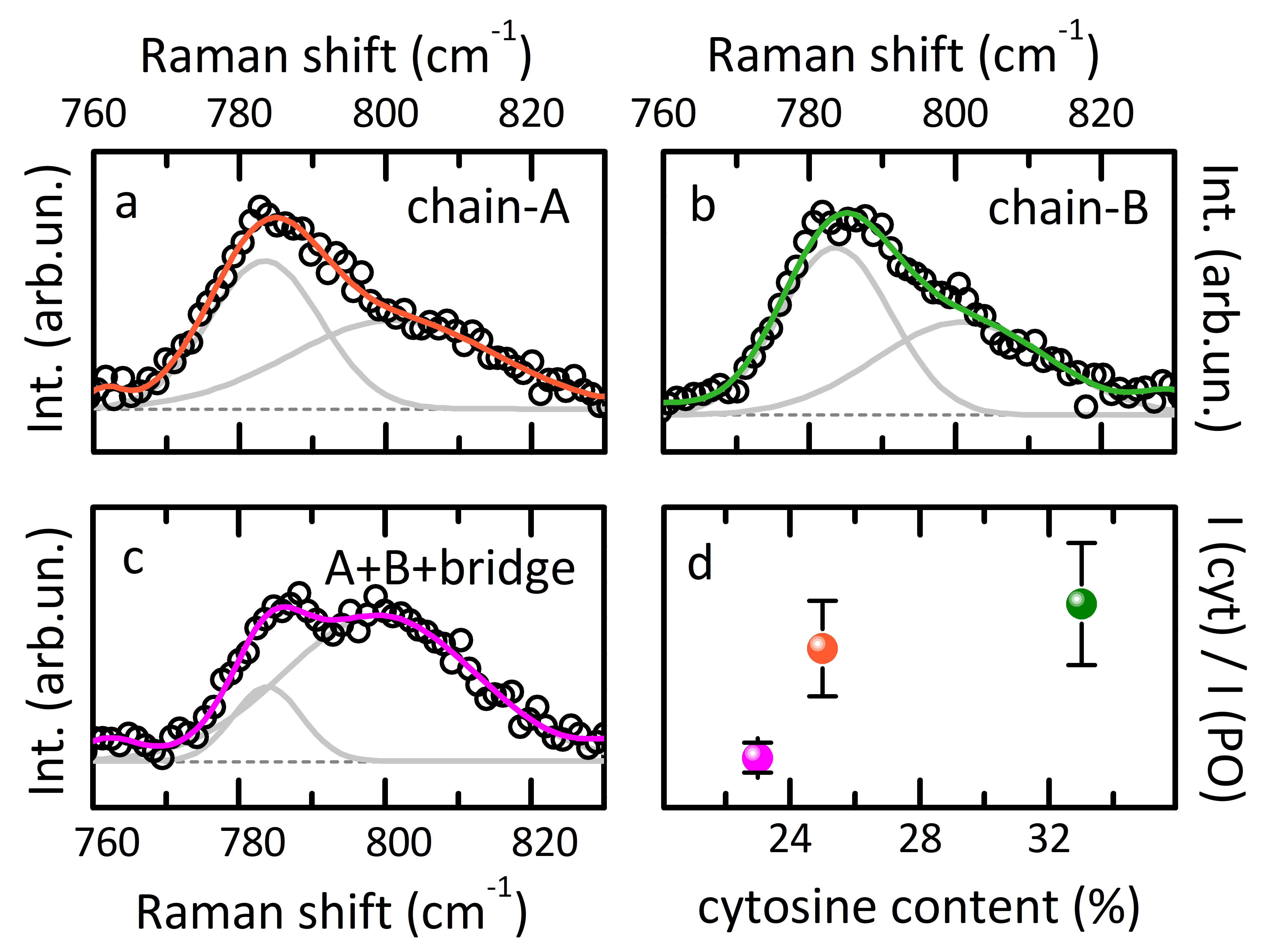}
\caption{\footnotesize Gaussian deconvolution of the SERS spectra of chain-A (a), chain-B (b) and bridge hybridized sample (c) in the 760-830 cm $^{-1}$ spectral region, where the band assigned to the cytosine and O-P-O vibrations is visible. In panel d, the ratio between the intensity of the low to high frequency components of the band is correlated with the concentration of cytosine in the probed sequences. } \label{figure-fit}
\end{figure}

As discussed above, a certain variability is expected to affect the high-frequency side of the spectra in Figure \ref{SERS_comparison_all} as a consequence of molecular orientation. On the other hand, a close inspection of the SERS band used for spectral normalization, around 784 cm$^{-1}$, reveals a modified shape for the bridge-conjugated system, compared to the single-chain spectra. It is generally accepted that this band originates from two distinct contributions, one ascribed to the cytosine mode, the other to the phosphate group \cite{thomas1995polarized, guerrini2015, guerrini2016, binren2015}. In particular, the inspection of the results in Refs. \cite{thomas1995polarized,guerrini2015} supports ascribing the low frequency component of the band to the cytosine vibration, and the higher component to the O-P-O stretching mode of the phosphate backbone. To quantify the spectral modification in the three spectra of Figure \ref{SERS_comparison_all}, we carried out a careful data fitting, deconvolving the contributions to the band into two gaussian functions, centered respectively at $783 \pm 1$ cm$^{-1}$ and $800 \pm 1$ cm$^{-1}$. The ratio between the intensity of the low frequency (cytosine) and high frequency (O-P-O) components nicely correlates with the amount of cytosine in the probed strands (Figure \ref{figure-fit} d), thus demonstrating that, despite the discussed sources of variability in the SERS spectral shape from sample to sample, the obtained SERS spectra reflect the DNA composition. The peak discussed in Fig. \ref{figure-fit} is far from the typical SERS fingerprint spectral region. Envisioning the use of this system as a biosensor, this will simultaneously enable to detect analytes and to monitor the state and composition of the molecular scaffold that holds the plasmonic nanosystem together. 

An aspect to be addressed is the less favorable signal-to-noise ratio that characterizes the spectrum of the A+B+bridge conjugated system compared to the spectra of the single-chain samples. This is clearly visible in Figure S6 of the Supplementary Material, where the raw data collected on single-chain and bridge conjugated system are presented along with the smoothed data of Figure \ref{SERS_comparison_all}. This is consistent with the different orientation of the DNA molecules in the two samples: for single-stranded functionalized NPs, the higher flexibility of DNA with respect to the double-stranded case can result in a more compact arrangement of the molecules in the proximity of the gold surface, where the enhancement is stronger, yielding to a clearer signal. 
Moreover, since most of the SERS signal arises from interparticle hot-spots, differences can be found in accidental and controlled NP aggregations (i.e. single and double strands). 

Support to the hypothesis of a controlled molecular arrangement, orthogonal to the NP surface, in the bridge mediated aggregates is given by the observation that a more pronounced reproducibility in the spectral intensity is retrieved on the bridge-conjugated system. As a representative example, we show in Figure S7 the most and least intense spectra acquired on a micrometric NP aggregate on the single-chain and bridge-combined samples. 
To investigate in detail the relation between the SERS response and the NP aggregation, we carried out high resolution FESEM imaging on both the single-chain AuNP depositions and on the bridge-conjugated aggregates.

\begin{figure}[]
\centering
\includegraphics[width=7.8cm]{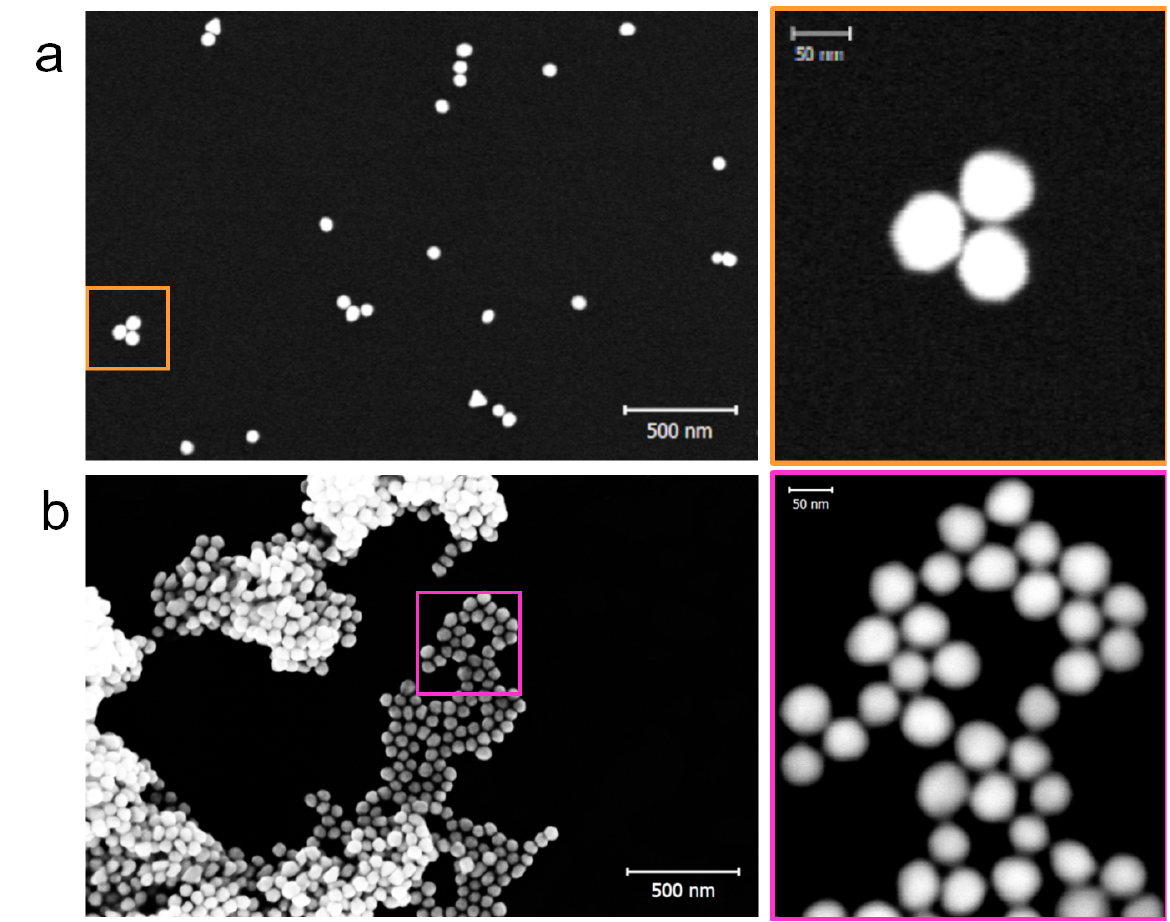}
\caption{\footnotesize Representative FESEM images of the SERS substrates for the single chain-A (a) and combined system A+B+bridge (b). For each sample a high resolution zoom of the highlighted detail is also shown.} \label{SEM_bridge}
\end{figure}

Representative results of FESEM microscopy are shown in Figure \ref{SEM_bridge}, where the images acquired on chain-A (panel a) and the A+B+bridge (panel b) samples are compared. The morphology of the SERS substrates appears clearly different in the two cases. In the single-stranded case, AuNPs are sparsely and randomly distributed onto the silicon substrate and only few small aggregates (dimers or trimers) can be found. Conversely, for the bridge-conjugated system AuNPs appear organized in large clusters. The zoomed images next to each frame show that the interparticle distance in the chain-A trimer is smaller than the hot-spot size within the A+B+bridge clusters. Remarkably, in the bridge conjugated substrate it is possible to recognize a junction between packed AuNPs that can be attributed to the bridge-connected double-stranded DNA chains.

For quantitatively clarifying the role of the nanoarchitecture in the originated SERS intensity and spectral quality, we analyzed the FESEM images in terms of radial autocorrelation function to extrapolate the average interparticle distance, as described in Section H of the Supplementary Material.
The analyzed data, reported in Table \ref{table_SERS}, reveals an average interparticle distance that is almost doubled in the  bridge-conjugated aggregates with respect to single-chain functionalized AuNPs. This depends on the semirigid, double helix DNA scaffold holding together the NPs when their aggregation is mediated by the bridge molecule \cite{DNAlength}. On the other hand, when the single-chain functionalized NP dispersion is deposited on a glass slide, the aggregation occurs by the uncontrolled self-assembly of the NPs, leading to smaller interparticle distances and more efficient, yet less reproducible hot-spots. Consistently, the lower signal revealed in the bridge-conjugated SERS system was accompanied by a higher reproducibility in the overall spectral intensity.  

\begin{table}[]
\caption{Inferred hot-spot size within aggregates}\label{table_SERS}
\centering
\begin{tabular}{lc}
\hline
sample & interparticle distance (nm)\\
\hline
chain-A & 2.6 $\pm$ 1.1 \\
chain-B & 3.0 $\pm$ 0.6 \\
A+B+bridge & 5.2 $\pm$ 0.6 \\
\hline
\end{tabular}
\end{table}

\section{Conclusions}

In conclusion, we proposed a stable, biocompatible SERS active system based on gold NPs covalently functionalized with specific DNA strands. The advantage of covalent DNA anchoring on the NP surface is twofold: on one hand, we demonstrated that this allows obtaining SERS spectra with reproducible and controlled spectral shapes, proving that the covalent functionalization prevents the accidental physisorption of different, random molecules on the NP surface \cite{colloids2016}. On the other hand, the covalent functionalization grants a well-defined orientation of the conjugated DNA strands on the NP surface, thus reducing variabilities in the SERS spectral shape \cite{marotta2013limitations} and sterically favoring the DNA hybridization. After functionalizing 60 nm gold NPs in a buffer solution with two purposely-selected 12-base single-stranded DNA sequences, we induced their aggregation into mesoscopic aggregates by adding to the NP dispersion a complementary, 24-base bridge sequence. We monitored the aggregation of the system as a function of temperature by UV-Visible absorption spectroscopy, proving the specificity of the DNA mediated interparticle pairing.  

The main features revealed by the DNA-SERS spectra cover a spectral window that is not superimposed to the usual fingerprint region of most of the analytes revealed in SERS biosensors \cite{xu2011sers, domenici2011, capocefalo2019}. This, along with the overall low Raman/SERS cross-section of DNA molecules, grants a good contrast for the potential spectroscopic detection of analytes in such a DNA-SERS biosensor. In the present study, the focus was not explicitly on the application of the system as a spectroscopic sensor: instead, we paid more attention to the thorough characterization of the nanosystems from both the composition and the structural point of view. In this respect, all the characteristics of the SERS spectra obtained from the DNA-conjugated nanosystems were justified and interpreted in terms of the system nanoscale architecture.

The spectral shapes in all the systems are very well-defined, as confirmed by repeated measurements, whereas the same consideration does not hold for the SERS intensity. We revealed a higher but less reproducible intensity in the single-chain functionalized systems, while the bridge-conjugated aggregated system features a more pronounced reproducibility. This is ascribed to the NP aggregation process, which is spontaneous and uncontrolled in the single-chain cases, leading to rare and more efficient hot-spots, while is well predictable and reproducible when the NP aggregation is driven by the specific pairing induced by the bridge molecule. The lower signal-to-noise ratio in the spectra of the bridge-conjugated system is due to the NPs being kept 5.2 nm far apart by the bridge linking. The results suggest that selecting shorter sequences for such a system can increase the intensity of the DNA-SERS spectra. The spectral shapes have been interpreted in terms of both molecular orientation and DNA strand base composition. 

We envision the application of the present system as a stable, versatile biosensor. The advantages of such a system can be numerous, among which, we recall the possibility to induce the (reversible) DNA mediated aggregation in presence of analytes by monitoring the temperature of the NP solution. Furthermore, by selecting the proper DNA strands, one can tune the interparticle distance, not only controlling the SERS enhancement but also favoring the trapping and sensing of smaller or bigger molecules. Finally, the base-pairing specificity can also be exploited for specific, targeted gene detection. 

\section*{Funding sources}
The authors acknowledge the Physics Department of Sapienza University for providing access to the CNIS facilities for FESEM measurements. 

\section*{Acknowledgements}
The authors would like to thank Dr. Francesco Mura for the technical support in FESEM imaging. D.C. acknowledges Sapienza University for funding "Avvio alla Ricerca" project 2017. 

\section*{Conflicts of interest}
The authors have no competing interests to declare.

\section*{Appendix A. Supplementary data}
Supplementary material related to this article can be found in the online version. 

\section*{References}

\bibliography{ms}

\end{document}